\documentclass[prd,floats,preprint,superscriptaddress,floatfix,nofootinbib,12pt]{revtex4}
\pdfoutput=1
\usepackage{amsmath}
\usepackage{graphics, graphicx}
\usepackage{epsfig}
\usepackage{ulem}
\usepackage{color}

\renewcommand{\Im}{{\rm Im}}

\def\bh{\hat a}

\def\Sh{\check S}

\def\Ih{\check I}

\def\1p{{(1p)}}
\def\be{\begin{equation}}
\def\ee{\end{equation}}
\def\beq{\begin{eqnarray}}
\def\eeq{\end{eqnarray}}
\def\p0{\phi_0}
\def\z0{\zeta_0}

\def\3G{^3{\cal G}}

\def\bh{{\hat b}}

\def\ol2{\frac{1}{\ell^2}}
\def\cc{k}

\def\df{}

\def\gf{}
\def\hf{}
\def\jf{}
\def\kf{}
\def\lf{}
\def\jf{}
\def\qf{}
\def\kf{}
\def\zf{}

\newcommand{\ttle}[1]{{\it #1}}

\begin{document}

\title{{\lf Quantum Probabilities for Inflation from Holography}}

\author{James B.  Hartle}
\affiliation{Department of Physics, University of California, Santa Barbara,  93106, USA}
\author{S.W. Hawking}
\affiliation{DAMTP, CMS, Wilberforce Road, CB3 0WA Cambridge, UK}
\author{Thomas Hertog}
\affiliation{Institute for Theoretical Physics, KU Leuven, 3001 Leuven, Belgium {\it and}\\
International Solvay Institutes, Boulevard du Triomphe, ULB, 1050 Brussels, Belgium}

\bibliographystyle{unsrt}

\begin{abstract}
The evolution of the universe is determined by its quantum state. The wave function of the universe obeys the constraints of general relativity and in particular the Wheeler-DeWitt equation (WDWE). For non-zero $\Lambda$, we show that solutions of the WDWE at large volume have two domains in which geometries and fields are asymptotically real. In one the histories are Euclidean asymptotically anti-de Sitter, in the other they are Lorentzian asymptotically classical de Sitter. {\zf Further, the universal complex semiclassical asymptotic structure of solutions of the WDWE implies that the leading order in $\hbar$ quantum probabilities for classical, asymptotically de Sitter histories can be obtained from the action of asymptotically anti-de Sitter configurations}. This leads to a promising, {\zf universal} connection between quantum cosmology and holography.

\end{abstract}



\maketitle

\section{Introduction}

Our large scale observations of the universe are of its classical behavior. The isotropic accelerated expansion and the large scale structure in the CMB and the galaxy distribution are just two examples.  {\jf  But it is an inescapable inference from the rest of physics that such classical features are governed by quantum mechanical laws.} A quantum system behaves classically when the probabilities implied by the quantum state are high for coarse-grained histories with correlations in time governed by deterministic equations of motion. That is true whether the system is a tennis ball in flight or the evolving universe as a whole. 
 
The form of the emergent deterministic equations of motion may be only distantly related to the equations defining the underlying quantum theory. {\lf Here we investigate this in theories with a non-zero cosmological constant $\Lambda$. In {\qf the usual formulation of} classical cosmology there is a sharp distinction between dynamical theories with different signs of $\Lambda$. The equations of a positive $\Lambda$ theory predict asymptotically deSitter (dS) geometries, whereas the equations of a negative $\Lambda$ theory predict asymptotically anti-deSitter (AdS) geometries.  

Quantum cosmology however is most fundamentally concerned with the prediction of {\it probabilities} for alternative histories of the universe. Those probabilities are derived from a theory of the universe's quantum state. In this paper we show that, in a natural formulation of quantum cosmology, the quantum probabilities obtained from a general wave function defined in terms of an asymptotically AdS action {\jf generally}  imply an ensemble of histories that, besides AdS  histories, also {\jf includes}  histories behaving classically according to equations of motion with a positive effective cosmological constant. 

{\jf Thus,} there is no sharp distinction between positive and negative $\Lambda$ in quantum cosmology. Instead the wave function of the universe naturally provides a unified framework in which both AdS and dS histories occur. 

\section{Classical Prediction in Quantum Cosmology}

A quantum state of a closed universe is represented by a wave function $\Psi$ on the superspace of three-geometries and matter fields on a closed spacelike surface $\Sigma$. {\jf For illustration} we consider a minisuperspace model\footnote{Our results are not restricted to minisuperspace models. The general situation is discussed in \cite{HHH12}.} with a negative cosmological constant $\Lambda$ (in the usual classical sense) and a single scalar matter field $\phi$ with potential $V=(1/2)m^2 \phi^2$. We take $m^2$ to be negative but within the Breitenlohner-Freedman (BF) range $m^2_{BF} < m^2 <0$, where $m^2_{BF} = -9/(4\ell^2)$ with $1/{\ell^2}\equiv -\Lambda/3$. The geometries are restricted to be spatially homogeneous, isotropic and closed\footnote{The generalization of our results  to open universes and universes with different topologies will be considered elsewhere.}.
{\gf The spatial geometries on a three-sphere can then be characterized by a scale factor $b$ and a boundary value $\chi$ of the homogeneous field $\phi$.} Three metrics describing possible three geometries are given by
\be
\label{threemetric}
d\Sigma^2 = b^2 d\Omega_3^2 .
\ee
A four-geometry is described by a one-parameter sequence of such scale factors. Cosmological wave functions are functions of $b$ and $\chi$, $\Psi=\Psi(b,\chi)$. 

{\lf A wave function predicts classical behavior in regions of superspace where} it is well approximated by a semiclassical form (or a sum of such forms) \cite{HHH08,HHH08b}
\be
\label{scform}
\Psi (b,\chi) = A(b,\chi) \exp(iS(b,\chi)/\hbar)
\ee
where $S$ varies rapidly compared to $A$. The classical histories are the integral curves of $S$. These are the solutions to the Hamilton-Jacobi relations $p=\nabla S$ relating the momenta involving time derivatives of the variables to the derivatives of $S$. The probability of a classical history passing through $(b,\chi)$ is proportional to $|A(b,\chi)|^2$.

\section{Asymptotic Structure of the Wheeler-DeWitt Equation}

Cosmological wave functions must satisfy the operator forms of the constraints of general relativity. The three momentum constraints are satisfied automatically as a consequence of the symmetries of this simple model.  The operator form of the remaining Hamiltonian constraint is the Wheeler-DeWitt equation (WDWE). With the simplest operator ordering this is\footnote{We have rescaled the scalar field from its usual value by a factor $\sqrt{(4\pi/3)}$ to simplify the equations.} 
\be
\left(-\frac{\hbar^2}{\eta^2}\frac{d^2}{d b^2} +\frac{\hbar^2}{\eta^2b^2} \frac{d^2}{d\chi^2}+ b^2 +\ol2 b^4-m^2 \chi^2 b^4\right)\Psi = 0  . 
\label{mini-wdw}
\ee
where $\eta \equiv 3\pi/2$. Defining $\Ih$ by
\be 
\Psi(b,\chi)\equiv \exp[-\eta\Ih(b,\chi)/\hbar] \ 
\label{defI}
\ee
the WDWE  becomes a non-linear equation for $\Ih$. In certain regions of superspace the terms that explicitly  involve $\hbar$ may be negligible compared to those that don't. When this semiclassicality condition is satisfied 
\be
- \left(\frac{d \Ih}{d b}\right)^2 +\frac{1}{b^2} \left(\frac{d \Ih}{d \chi}\right)^2 + b^2 +(\ol2-m^2 \chi^2) b^4 =0
\label{mini-HJ}
\ee
which is independent of $\hbar$. Substituting a solution of \eqref{mini-HJ} into \eqref{defI} gives a leading order in $\hbar$ semiclassical approximation to the wave function.

The semiclassicality condition is satisfied for sufficiently large $b$ provided $\Lambda\ne 0$. This can be established self consistently by assuming it holds, solving \eqref{mini-HJ} for large $b$ and checking that the neglected $\hbar$ terms in the WDWE are negligible
\cite{HHH12}. Asymptotically in $b$ {\df sufficiently} general solutions to the Hamilton-Jacobi equation \eqref{mini-HJ} are defined by the expansion
\be
\Ih(b,\chi)=\frac{\cc_3}{\ell} b^3 + \cc_2 b^2+\frac{\ell}{2} \cc_1 b +\cc_{-}b^{\lambda_{-}} + \cc_0+\cdots 
\label{asymptact}
\ee
where the $\cc$'s are all functions of $\chi$ and $\lambda_{\pm} \equiv  (3/2)(1\pm\sqrt{1+4\ell^2 m^2/9})$.
Substituting this in \eqref{mini-HJ} yields the following equations that determine the leading coefficient functions,
\begin{subequations}
\label{ceqns}
\be
\label{c3}
-9\cc_3^2+(\cc_3')^2 +1-\ell^2 m^2 \chi^2 =0 ,
\ee
\be
\label{c2}
-6\cc_3\cc_2+\cc_3' \cc_2' =0, 
\ee
\be
\label{c1}
-3\cc_3\cc_1+\cc_3' \cc_1'-4\cc_2^2+(\cc_2')^2+1 =0,
\ee
\be
\label{c0}
-3\lambda_{-} \cc_3\cc_{-}+\cc'_3\cc'_{-}=0 ,
\ee
\be
\label{c4}
(2/\ell)\cc_0' \cc_3' +\cc_2'\cc_1'-2\cc_2\cc_1=0 ,
\ee
\end{subequations}
a prime denoting the derivative with respect to $\chi$.  

The first three of these equations are a closed set for the first three coefficients in \eqref{asymptact}.
Were there no field they would have the unique solution $\cc_3=1/3$, $\cc_2=0$, and $\cc_1=1$. We use these values as boundary conditions to fix a unique solution for the leading terms at small $\chi$ of the first three coefficients. The small $\chi$ behavior of these terms is therefore universal --- the same for all wave functions satisfying the WDWE.  
By contrast, the large $\chi$ behavior as well as the remaining terms in \eqref{asymptact} depend on the specific wave function. 

From \eqref{asymptact} it follows that the semiclassicality condition holds at large $b$ as assumed, and that the general solution $\Psi(b,\chi)$ of the WDWE has the asymptotic form 
\be
\label{wavegen}
A_{+} \exp[-\eta \Ih( b,\chi)/\hbar] + A_{-} \exp[+\eta \Ih (b,\chi)/\hbar] 
\ee
for arbitrary constants $A_\pm$.

{\df Any solution of \eqref{mini-HJ} with a sufficient number of arbitrary constants will determine a class of solutions to the equations of motion of general form. Specifically, sufficiently general solutions of the Hamilton-Jacobi equation \eqref{mini-HJ} imply general first integrals of the Einstein equations. Hence} the expansion \eqref{asymptact} encodes the Fefferman-Graham (FG) asymptotic expansion of solutions to the Einstein equations with a negative cosmological constant \cite{Fefferman85}. The universal behavior at small $\chi$ of the first three terms in \eqref{asymptact} corresponds to the universal terms in the FG expansion.

\section{Two Real Domains}

{\lf To identify the ensemble of histories} predicted by \eqref{wavegen} we search for domains of asymptotic superspace that correspond to real three-geometries and field configurations. {\kf  The theory predicts classical evolution in domains where} the wave function is a sum of semiclassical terms like \eqref{scform}, with $S$ varying rapidly compared to $A$.}  {\qf The union of all  sets of classical histories predicted in such domains is  the ensemble of possible classical histories predicted by the wave function.}

There are two domains representing real three geometries corresponding to real values of $b$ and purely imaginary values $b=i\bh$, $\bh$ real \footnote{ {\kf The two domains are represented by metrics of opposite signature so considering both on an equal footing is neutral with respect to the choice of signature convention.}}.   

{\kf The specific case of the no-boundary wave function \cite{HH83} provides strong motivation for considering both domains on an equal footing in the process of classical prediction. This is because the NBWF's semiclassical approximation involves regular complex saddle points, which provide a very close connection between both domains in the wave function \cite{HH11}. Specifically we have shown \cite{HH11} that the semiclassical wave function in one domain can be written as a wave function defined in terms of saddle points associated with the second domain (up to universal surface terms). This implies in particular that the probabilities of the classical, asymptotically dS histories predicted by the NBWF are given by the regularized actions of asymptotically Euclidean AdS domain walls. Moreover this connection between both domains in the NBWF emerges automatically from its holographic formulation \cite{HH11}, where both domains correspond to complex deformations of a single underlying dual field theory.

These results strongly motivate treating both domains on an equal footing in the wave function by extending its configuration space appropriately. This yields a natural generalization of the usual framework for classical prediction in quantum cosmology in which one takes the classical ensemble predicted by any wave function of the universe to consist of the classical histories from both real domains. We now explore the implications of this for the ensemble of asymptotic histories in the simple models considered above.

\section{A Model Classical Ensemble}

To exhibit the histories explicitly we first restrict to the region of superspace where $\chi$ is small so the field can be treated perturbatively. The small $\chi$ solutions of \eqref{ceqns} for the first three, universal terms in \eqref{asymptact} are \footnote{Replacing $\lambda_{-}$ by $\lambda_{+}$ in the first equation of \eqref{sol1} also gives a perturbative solution but a numerical integration of the non-linear eq \eqref{c3} shows that \eqref{sol1} is the generic one.}:
\be
\label{sol1}
\cc_3  =\frac{1}{3}+\frac{1}{2} \lambda_{-} \chi^2, \quad \cc_2 =0, \quad \cc_1=1+\frac{3}{4}\chi^2.
\ee
The leading non-universal coefficients are given by
\be
\label{sol2}
\cc_{-}  =K \chi, \quad  \cc_0=J
\ee
where $K$ and $J$ are complex constants (independent of $\chi$). They are are not determined by the Hamilton-Jacobi equation but depend on the specific wave function. The relative probabilities of the predicted classical histories therefore depend on them \cite{HHH08}. 

Since $\cc_3$ in \eqref{sol1} is real no classical histories are predicted for the domain where $b$ is real because there $\Ih$ is approximately real and there is no imaginary part that varies rapidly. However, in the domain where $b=i\bh$ the wave function \eqref{wavegen} will be a sum of potentially semiclassical forms \eqref{scform} with $S=\eta\Sh$, $\Sh=-\Im \Ih$, and the leading order asymptotic behavior
\be
\label{wave2}
 \Sh(\hat b,\chi) = \frac{\hat b^3}{3\ell} + \frac{1}{2\ell} \lambda_{-} \chi^2\hat b^3+\cdots
\ee
which varies rapidly for large $\bh$ so that the wave function predicts an ensemble of classical histories.

The Lorentzian histories that comprise the classical ensemble are the integral curves of $\Sh$. The $\Sh$ above is, in fact the action of a homogeneous and isotropic, asymptotically de Sitter history with $\Lambda = 3/\ell^2$, perturbed by a scalar field moving in a quadratic potential with positive mass $-m^2$. 

The ensemble of asymptotic classical histories can be obtained explicitly by solving the Hamilton-Jacobi equations connecting the momenta to the gradients of $\Sh$ and defining the integral curves. For this simple model these equations are
\be 
\frac{d\bh}{dt} = \frac{1}{\bh} \frac{\partial\Sh}{\partial\bh},   \quad \frac{d\chi}{dt}=-\frac{1}{\bh^3} \frac{\partial\Sh}{\partial\chi} \ .
\label{HJeqns}
\ee
Substituting \eqref{sol1} and \eqref{sol2} in \eqref{asymptact} we find for the asymptotic scale factor, with $u \equiv  \exp(-t/\ell)$,
\be
\bh(t) = \frac{2\ell}{u}[1 + u^2-\frac{3}{4\alpha^2} u^{2\lambda_{-}}+\cdots ]
\label{accexpn}
\ee
and for the asymptotic scalar field \footnote{When $\lambda_{-}<1/2$ there is an additional universal term 
$ (-3\alpha/8) u^{2+\lambda_{-}}$.}
\be
\label{asympchi}
\chi(t) = \alpha u^{\lambda_{-}} +\beta u^{\lambda_{+}}+\cdots
\ee
where $\alpha$ is an integration constant of the first order equation for $\chi$ in \eqref{HJeqns} and  $\beta=-\Im(i^{\lambda_{-}} K)/(\lambda_{+}-\lambda_{-})$. Eqs \eqref{accexpn} and \eqref{asympchi} describe an ensemble of asymptotic deSitter spaces perturbed by a homogeneous decaying scalar field \cite{Starobinsky83}.

As $\ell$ approaches zero the lower bound on values of $b$ for which the asymptotic expansion \eqref{asymptact} holds become larger and larger. If the cosmological constant vanishes then there can still be classical histories (e.g. \cite{HHH08}) but their existence and character depends on the specific form of the theory and the wave function. If there is no matter then there are no classical histories when $\Lambda=0$. 

\section{Slow Roll Inflation}

Beyond the perturbative regime,  there is a solution of \eqref{ceqns} which at large $\chi$ behaves as 
\be
\label{slo}
\cc_3 (\chi) \approx \frac{\ell m}{3}\chi,  \quad \cc_2 =0,  \quad \cc_1\approx 1/\ell m\chi
\ee
and at small $\chi$ tends to \eqref{sol1}. This solution for $\cc_3$ is shown in Fig \ref{cfunct}. 
The leading contribution to the asymptotic action in the large $\chi$ regime is then
\be 
\label{wavu4}
\Sh(\hat b,\chi)\sim \frac{m\chi \hat b^3}{3} + \cdots. 
\ee
which is characteristic of cosmologies undergoing slow roll scalar field inflation \cite{HHH08,Lyo92}. Indeed the leading order solutions of \eqref{HJeqns} derived from \eqref{wavu4} exhibit the familiar slow roll behavior, $\chi(t)=\chi(0)-mt/3$. 

This shows that the presence of a regime of classical slow roll inflation can be deduced from the semiclassical asymptotic form of the wave function representing a quantum state. That could be useful in a holographic study of which quantum states predict slow roll inflation.

\begin{figure}[t]
\includegraphics[width=3.0in]{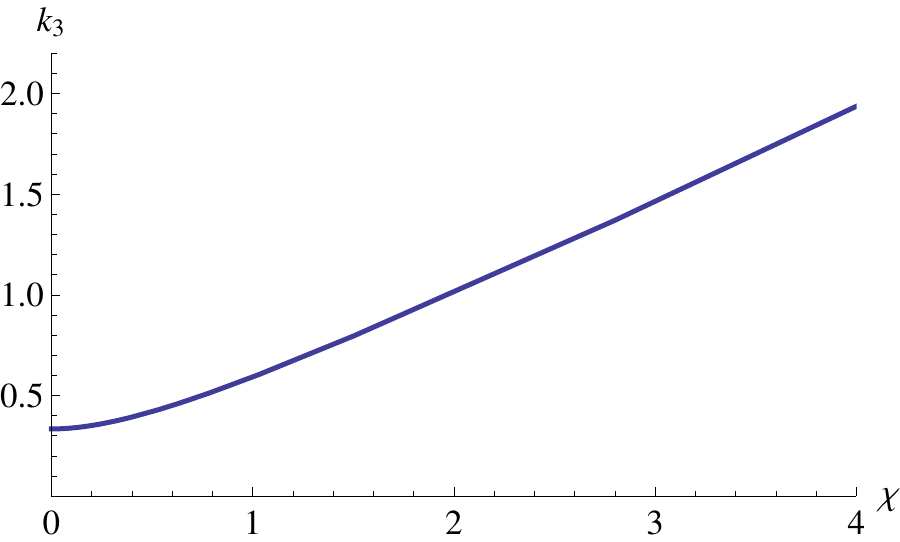} 
\caption{A solution of \eqref{c3} for the leading, universal coefficient function $\cc_3$ that specifies the asymptotic wave function. The approximately linear behavior for large $\chi$ is characteristic of field driven, slow roll inflation.}
\label{cfunct}
 \end{figure}

\section{Holographic Probabilities}

{\lf The central objective of quantum cosmology is to compute {\it probabilities} for alternative classical histories of the universe.
To leading order in $\hbar$ these are given by the asymptotically finite real part of the action \eqref{asymptact} in a regime where the wave function predicts classical behavior. 

The real contributions to \eqref{asymptact} depend on the boundary conditions implied by a specific theory of the wave function. 
However the asymptotic structure of solutions of the WDWE discussed above gives rise to a universal complex extension of the asymptotic Fefferman-Graham expansions which includes the Starobinsky expansion \cite{Starobinsky83} of asymptotic dS solutions. Using these expansions we recently showed that, for any wave function, the asymptotic action of a general solution in the dS domain can be written in terms of an asymptotic AdS action and a number of universal `surface' terms that are fully determined by the argument of the wave function \cite{HH11}. This means that, at the level of the complex solutions specifying the wave function, both domains are closely interconnected. Specifically, the leading order probabilities for the classical, asymptotically de Sitter histories predicted by any wave function can be obtained from the action of Euclidean asymptotically AdS domain walls\footnote{The scalar profile along the AdS domain walls that enter in this analysis is in general complex as discussed in detail in \cite{HH11}.}.

This leads to a natural connection with the Euclidean AdS/CFT correspondence (see also 
\cite{Maldacena03,Horowitz04,HH11,Fadden10,Anninos11}). The AdS/CFT duality makes a distinction between the universal contributions to the asymptotic wave function, which grow as a function of the scale factor\footnote{In field theory applications of AdS/CFT these are treated as UV divergences and cancelled by counterterms.}, and the non-universal terms that provide the asymptotically finite contribution and govern the relative probabilities of different configurations. The duality conjectures that the latter are related to the partition function $Z$ of a dual Euclidean field theory defined on the boundary surface $\Sigma$. 

In the perturbative solution of the minisuperspace model given above the universal terms are given in \eqref{sol1} and the non-universal terms in \eqref{sol2}. In a regime where the low energy gravity approximation can be trusted the duality then states that, to leading order in $\hbar$ and at large volume,
\be\label{duality}
Z[\tilde \chi] = \exp (-\eta (K \tilde \chi+J)/\hbar) 
\ee
where  $J$ and $K$ are defined in \eqref{sol2} and  $Z$ is defined by 
\be\label{part}
Z[\tilde \chi] \equiv \langle \exp \int d^3x \sqrt{\tilde \gamma} \tilde \chi {\cal O} \rangle_{QFT}
\ee
with $\cal O$ the dual operator that couples to the source $\tilde \chi = b^{\lambda_{-}}\chi$ induced by the bulk scalar.
Differentiation of $Z$ with respect to $\tilde \chi$ yields $\langle  {\cal O} \rangle$ which provides an alternative way to calculate $K$ in \eqref{duality}.

The scalar argument of the wave function thus enters as an external source in the dual field theory that turns on a deformation. 
In field theory applications of AdS/CFT the value of $\tilde \chi$ is usually held fixed. By contrast, in a cosmological context one is interested in the wave function of the universe as a function of $\chi$. In this context the dependence of $Z$ on the value of $\tilde \chi$ yields a measure on different configurations on $\Sigma$ \cite{HH11}. 

Hence in cosmology one is led to consider all deformations induced of a given CFT that correspond to real boundary configurations and for which the partition function converges. In the toy model discussed here this means one should include purely real deformations, corresponding to real values of $b$, as well as deformations for which $\tilde \chi$ has the phase $\exp[i\lambda_{-}\pi/2]$. Together these generate the two real domains of asymptotic superspace discussed earlier.

In a dual formulation of the wave function, the difference between both domains thus shows up only in the phase of the scalar deformation of a single dual field theory. Since the phase of $\tilde \chi$ in general has an effect on the partition function this means that the asymptotic wave function will be a different function of $\chi$ in both domains. We illustrat{\hf ed} this with the example of the no-boundary wave function in \cite{HHH12}.

\section{Discussion}

We have used a natural framework for the prediction of probabilities of classical histories in quantum cosmology to show that accelerated expansion can be a low energy prediction of theories of the wave function of the universe {\lf defined in terms of what would classically be a negative cosmological constant. Specifically we have shown that asymptotically AdS wave functions satisfying the constraints of general relativity have a universal semiclassical asymptotic structure for large spatial volumes which implies} that the wave function describes both a set of asymptotically Euclidean AdS histories as well as an ensemble of histories which expand, driven by a positive effective cosmological constant. 

The relative probabilities of {\it both sets of histories} are given by the asymptotically AdS actions of (possibly complex) Euclidean AdS domain walls. This leads to a natural connection between quantum cosmology and the general framework of Euclidean AdS/CFT, not as a map from an AdS theory to a deSitter theory as envisioned in \cite{Maldacena03,Fadden10,Anninos11} but within the context of a single theory of the wave function of the universe.
 
{\lf At the level of the histories emerging from the wave function one could argue our results imply that the sharp distinction between positive and negative $\Lambda$ in classical physics {\qf as it is usually formulated} disappears in quantum cosmology. Rather the wave function provides a unified arena encompassing both kinds of histories. 

The generalization of these results to include inhomogeneities and {\lf the prospects of embedding the models considered here in fundamental theory} are described in \cite{HHH12}. 

\noindent{\bf Acknowledgments:} JH and SWH thank Marc Henneaux and the International Solvay Institutes for their hospitality. The work of JH was supported in part by the US NSF grant PHY08-55415. The work of TH was supported in part by the National Science Foundation of Belgium under the FWO Odysseus program.

\end{document}